# Precision of Evaluation Methods in White Light Interferometry: Correlogram Correlation Method

Ilia Kiselev,[1,*] Egor Kiselev,[2] Michael Drexel,[1] Michael Hauptmannl[1]

[1]Breitmeier Messtechnik GmbH, Englerstr. 24, Ettlingen, D-76275, Germany
[2]Institute for Theoretical Condensed Matter physics, Karlsruhe Institute of Technology, Karlsruhe D-76131, Germany
*Corresponding author: kiselev@breitmeier.de

## Abstract

In this paper we promote a method for the evaluation of a surface's topography which we call the correlogram correlation method. Employing a theoretical analysis as well as numerical simulations the method is proven to be the most accurate among available evaluation algorithms in the common case of uncorrelated noise. Examples illustrate the superiority of the correlogram correlation method over the common envelope and phase methods.

**Key words:** Surface topography, coherence scanning interferometry, noise stability

## Introduction

From the early times of the White Light Interferometry (WLI) it has been clear that WLI is a powerful tool to determine the topography of a surface [1]: The WLI signal is a wave packet *I*, and the shift of its position $z_0$ on the scanning axis $z$ identifies a change of local height of the reflecting surface. There exist two established methods to localize the $z_0$ on the axis $z$ [2]: the phase method stemming from the monochrome interferometry (PSI - Phase Shifting Interferometry) and the correlogram envelope evaluation method (CSI - Coherence Scanning Interferometry) which relies on the properties of the broad WLI signal spectrum. However, both methods harvest only parts of the information contained in measured correlogram. As a consequence, the envelope evaluation methods suffer from low precision [2] and PSI is subjected to the 2π ambiguity of the phase determination [3]. Numerous attempts exist to marry both methods [4], [5], but their success is limited, because the correlogram's information is still only partly employed. To use the complete information one has to consider the full shape of the correlogram In the absence of noise/other disturbances - the only changes in the shape of the correlogram are contrast scaling and a shift of its position along the scanning axis. In order to locate the surface one has just to search for the expected wave package pattern on the scanning axis and the best way to do so is to find the position of maximal cross-correlation with a reference corellogram. The arising method, which we name correlogram correlation or, for short, CorCor method, is illustrated in Fig. 1.

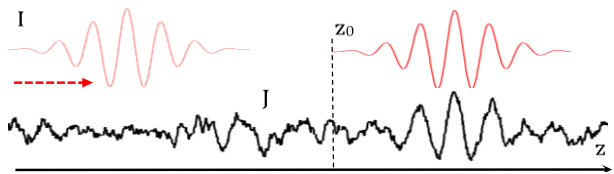

Fig. 1. CorCor method: search for $z_0$ – the best fitting position of the reference correlogram *I* on a measured correlogram *J*.

Our intention here is to promote this cross-correlation method to a due level of application; to this purpose we prove that the method is necessarily optimal in the sense of its precision in the presence of noise, except for rare, specific disturbances. An additional important advantage of the method is the fact, that it provides a direct criterion for the appropriateness of a measured local correlogram - the covariance with the reference correlogram at the best fitting position. This criterion is more informative and useful than the commonly employed criterion of the magnitude of maximal contrast. As far as the non-normalized covariance is used, this criterion also includes the magnitude of the local correlogram; however, in this paper this advantage is not dwelled on.

Of course, the basic idea to look for the position of the wave packet in the course of measured correlogram has been touched on earlier, e.g. in [6-8], but to our best knowledge it is still ignored in daily applications. This is because it has not been realized until now that this method is the superior to other procedures in precision and stability in a strict mathematical sense. The advantage is rooted in the full use of available information; in combination with the maximal-cross-correlation technique the superiority can be easily shown mathematically. Previously, the method was just considered as an additional specific possibility: in this way, the cited works [6], [7] apply it to the special case of transparent film metrology. While [7] employs a complicated window shifting procedure to find the wave package position, [8] already suggests to use the cross-correlation technique. However, the authors of [8] do not consider the idea of a reference correlogram, but use a model wave packet instead. Since the model wave packet's shape differs from the measured one, the information is lost again - the issuing precision of the method deteriorates.

## Results and discussion

### 1. Establishing the CorCor method

Let us obtain the surface height $z_0$ following the Maximal Likelihood Criterion (MLC) [9]. Let $I_j \equiv I(z_j - z_0)$ be the reference correlogram of the interferometer shifted to the position $z_0$, where $z_j$ are measurement points on the z-axis. Then for the measured correlogram $J_j$ holds

$$J_j = I_j(z_0) + \delta_j, \tag{1}$$

where $\delta_j$ represents a discrepancy between $J_j$ and the reference correlogram. The discrepancy appears due to both z-shift of correlogram caused by the varying height of the reflecting surface and noise, which we assume to be uncorrelated and Gaussian. We imply that the mean levels of all correlograms are removed before the height processing. Following the MLC, among possible shift positions $z_0$ of $I_j$ we have to choose the one, at which the probability that the measured correlogram is constituted by the $I_j$ and noise $\delta_j$ is maximal. This probability is given by

$$p = \prod_{j=1}^{N} p_{\delta_j} = \left(\frac{1}{\sqrt{2\pi}\sigma}\right)^N \exp\left(-\sum_{j=1}^{N} \delta_j^2 / 2\sigma^2\right), \tag{2}$$

were $N$ is the number of measurement points, $p_{\delta j}$ are pointwise probabilities, and $\sigma^2$ is the noise dispersion supposed to be equal for all the points. The maximization of (2) automatically results in the least-square requirement:

$$\sum_j \delta_j^2 \to \min, \quad \text{or} \quad \sum_j \left(J_j^2 - 2J_j I_j + I_j^2\right) \to \min, \tag{3}$$

which is equivalent to the requirement of maximization of covariance:

$$\sum_j J_j I_j(z_0) \to \max_{z_0}. \tag{4}$$

Thus, the most probable shift of the correlogram and hence the most probable surface position is at a $z_0$, where the covariance of the measured and reference correlograms is maximal. In other words, to be in accordance with the MLC and to get the surface position one has to calculate the cross-correlation function and to find the position of its maximum:

$$(J \otimes I)(z_0) \to \max_{z_0}. \tag{5}$$

This is the correlogram correlation (CorCor) method. Let us emphasize: according to the above derivation no other procedure can give a more accurate estimation of surface height in the sense of its probability. Hence, the robustness to noise of this estimation procedure cannot be surpassed [9]. Note that this statement is also correct for high noise amplitudes as for low.

The rare net of measurement points $z_j$ typically used in WLI, does not mean that the maximum in (4) or (5) can be found only at one of these discrete points. It means instead that harmonics with frequencies above the Nyquist limit are not present in correlograms. This limitation does not prevent us from finding the maximal correlation position exactly. In this study we have calculated the cross-correlation function (4) on the rough net of measurement points and then interpolated it using an interpolation method which preserves the spectrum of interpolated function. This procedure is equivalent to the interpolation of correlograms and following calculation of their cross-correlation function. It is, however, much faster[1].

## 2. Cramon-Rao estimation for the CorCor method

Now we are going to obtain analytical expressions for the noise variance of the CorCor estimation of $z_0$, that of the established methods, and will compare them. Suppose, there is a sample set of values $J_j$ (the correlogram) which serves as basis for the estimation of parameter $z_0$. The Cramer-Rao bound [10] gives the lowest possible estimation of this parameter's variance. If the probability for the appearance of the set is known, and in our case it is given by expression (2), the variance, according to the Cramer-Rao bound satisfies the inequality:

$$\text{var}(z_0) \geq -1 / E\left(\partial^2 \ln p / \partial z_0^2\right), \quad (6)$$

where $E$ stands for taking the statistical expectation. As it will be demonstrated by simulation below, in (6) the equality is actually in effect, i.e. the Cramer-Rao estimation is efficient - the lower variance bound is reached by the method. If (4) is employed to get the value of $z_0$ for the sample set $J_j$, the (1) and (2) hold, and the logarithm of (6) and its second derivative can be written as

$$\ln p = -\sum_j \delta_j^2 / 2\sigma^2 + C;$$

$$\frac{\partial^2 \ln p}{\partial z_0^2} = \frac{\partial^2}{\partial z_0}\left[-\frac{1}{2\sigma^2}\sum_j -2\delta_j \frac{\partial I_j(z_0)}{\partial z_0}\right] = \frac{1}{\sigma^2}\sum_j\left\{-\left[\frac{\partial I_j(z_0)}{\partial z_0}\right]^2 + \delta_j \frac{\partial^2 I_j(z_0)}{\partial z_0^2}\right\};$$

where C is a constant and we have taken into account that the deviations $\delta_j$ do not depend on $z_0$. Furthermore, only the $\delta_j$ are variates, and they have zero mean values, thus

$$E\left(\delta_j \cdot \partial^2 I_j(z_0) / \partial z_0^2\right) = 0,$$

and, finally,

$$\text{var}(z_0) = \sigma^2 \Big/ E\left(\sum_j\left\{\left[\frac{\partial I_j(z_0)}{\partial z_0}\right]^2 - \delta_j \frac{\partial^2 I_j(z_0)}{\partial z_0^2}\right\}\right) = \sigma^2 \Big/ \sum_j\left[\frac{\partial I_j(z_0)}{\partial z_0}\right]^2. \quad (7)$$

This result agrees with the common sense expectation that the positioning is exacter for steeper shapes of the pattern to be fitted. We want to emphasize that, just as (5), the estimation (7) is valid for noises of any magnitude.

## 3. Cramer-Rao estimation for the CSI methods. Envelope-parabola method

---

[1] To be published.

The usual way to get the envelope $\hat{E}(z)$ (often referred as the Hilbert transform method) is the following. The reference correlogram located at the position $z_0 = 0$ can be written as (see, e.g., [11])

$$I(z) = \int_0^\infty \Psi(k)\cos(2kz)dk. \tag{8}$$

We can construct the conjugate function

$$I_s(z) = \int_0^\infty \Psi(k)\sin(2kz)dk, \tag{9}$$

which is the Hilbert transform of $I$:

$$I_s(z) = 1/\pi \int_{-\infty}^\infty I(z')/(z-z')dz'$$

and then obtain the envelope as

$$\hat{E}_0(z) = \sqrt{I^2(z) + I_s^2(z)}. \tag{10}$$

Similarly, for a noisy correlogram $J_j = I_j + \delta_j$, we obtain $J_{sj} = I_{sj} + \delta_{sj}$ and, taking in account that the $\delta$-s are small, we write

$$\hat{E}_j = \sqrt{J^2_j + J^2_{sj}} \approx \hat{E}_{0j} + \frac{I_j}{\hat{E}_{0j}}\delta_j + \frac{I_{sj}}{\hat{E}_{0j}}\delta_{sj}. \tag{11}$$

Here $\hat{E}_0$ is the envelope of the reference correlogram $I$. If $\sigma^2$ is the dispersion of the correlogram's deviation $\delta_j$ and, hence, also of $\delta_s$, the standard deviation of the envelope is

$$\sigma(\hat{E}_j) = \sqrt{\left(\frac{I_j}{\hat{E}_{0j}}\right)^2 \sigma^2 + \left(\frac{I_{sj}}{\hat{E}_{0j}}\right)^2 \sigma^2} = \sigma, \tag{12}$$

Therefore, the deviations of the envelope are also normal distributed variates of the same dispersion. In CSI one estimates the envelope maximum position or the position of one of its centroids and ascribes it to the local surface height. Both procedures are widely used [2]. In so doing one needs to ascertain a credible part of the envelope curve. Often it is done by extracting the envelope part, which lies above $E_{max}/2$, where $E_{max}$ is the maximal value of the envelope (10). We call this extracted part "half-height envelope". In this subsection we consider the envelope fitting with a parabola, whose maximum position is then taken as the location of envelope maximum. In the following subsection we turn to the centroid method.

A curve is fitted to the envelope using the least-square method. The most suitable is the fitting with the reference envelope $\hat{E}_0$ as the origin of $\hat{E}$. According to the discussion of subsection 1. this is equivalent to finding the position of best correlation. Thus, according to the subsection 2., the variance of $z_0$ obtained in this way can be estimated as

$$\text{var}(z_0) = \sigma^2 \Bigg/ \sum_j \left[\frac{\partial \hat{E}_{0j}(z_0)}{\partial z_0}\right]^2. \tag{13}$$

This is a variance estimation for the envelope-fitting methods, among others for the half-height envelope fitting with a parabola. It provides just the lower bound for the variance because of the following two reasons: i) The expression (13) corresponds to the least square (LS) fitting with a reference envelope, the fitting with a parabola can have higher deviations. ii) Although the envelope variations (11) are normally distributed with standard deviations (12), they are positively correlated with their neighbor deviations, therefore the probability has a more complex expression, than (2) (not derived here). The fact that ii) results in higher variations than the ones given by (13) is illustrated in the next subsection and confirmed by numerical simulation.

Since the envelope $\hat{E}_0$ in (13) is much less steep, than the correlogram $I$ in (7), the variance of surface height estimation for the envelope approach is much higher, than for the CorCor method, showing that the latter is more accurate.

## 4. Variance of height estimation by the envelope centroid method

An alternative procedure to obtain the envelope's position on the scanning axis is the center-of-gravity (centroid) estimation (see, e.g., [12]). Assuming the validity of (12) for the dispersion of deviations of the envelope due to noise we directly derive an estimate for the $z_0$ variance of this method. Let us here and below take the length of step on the discretized $z$ axis as the length unit, so that $\Delta z = 1$; $z_j = j$. In this subsection, if other is not specified, the summation index $j$ changes in limits of the half-height envelope. According to the first order centroid method, the surface position $z_0$ is determined using the following formula:

$$\tilde{z}_0 = \sum j \hat{E}_j \Big/ \sum \hat{E}_j . \tag{14}$$

If $z_0$ is the true position and $\hat{E}_0$ the envelope of the reference correlogram at this position, i.e. $z_0 = \Sigma j \hat{E}_0 / \Sigma \hat{E}_0$, then

$$\tilde{z}_0 = \frac{\sum j(\hat{E}_{0j} + \delta_j)}{\sum (\hat{E}_{0j} + \delta_j)} = \frac{\sum j \hat{E}_{0j} + \sum j \delta_j}{\sum \hat{E}_{0j}} \left( \frac{1}{1 + \sum \delta_j / \sum \hat{E}_{0j}} \right) =$$

$$\frac{\left( \sum j \hat{E}_{0j} + \sum j \delta_j \right) \left( \sum \hat{E}_{0j} - \sum \delta_j \right)}{\left( \sum \hat{E}_{0j} \right)^2} + o\left( \frac{\sum \delta_j}{\sum \hat{E}_{0j}} \right) =$$

$$= z_0 - z_0 \frac{\sum \delta_j}{\sum \hat{E}_{0j}} + \frac{\sum j \delta_j}{\sum \hat{E}_{0j}} + o\left( \frac{\sum \delta_j}{\sum \hat{E}_{0j}} \right),$$

and accurate to the infinitesimal of higher order the deviation of the height estimation due to noise is given by[1]

$$\delta_{z_0} = \tilde{z}_0 - z_0 = \sum (j - z_0) \delta_j \Big/ \sum \hat{E}_{0j} . \tag{15}$$

The envelope deviation distribution over $j$ has an autocorrelation function with finite width (proportional to the ratio of envelope width to main wave period) [13]. The $\delta_j$-s are not independent, but positively correlated with their neighbors. According to the property of variance of summed random variables [14], the variance of the $\delta_{z0}$ can be obtained from (15) as

$$\sigma_{z_0}^2 = \frac{\sigma^2 \sum_j (j-z_0)^2 + 2 \sum_j \sum_{k<j} (j-z_0)(k-z_0) K_{jk}}{\left( \sum \hat{E}_{0j} \right)^2}.$$

Neglecting the covariances $K_{jk}$, which is reasonable, because the white-light correlogram wave packets contain few main periods, we obtain a lower bound for the variance of $z_0$:

$$\text{var}(z_0) \equiv \sigma_{z_0}^2 \leq \sigma^2 \sum (j-z_0)^2 \Big/ \left( \sum \hat{E}_{0j} \right)^2 . \tag{16}$$

---

[1] Actually, (15) is not exact even by zero noise but should be corrected to eliminate systematic errors. Here, we consider only the noise-induced deviation of an unbiased height estimation, where systematic errors has already been corrected.

## 5. Variance of height estimation by the phase method

The variance of the height estimation following the correlogram phase evaluation method (PSI) is obtained in [15] and [16] by applying the Cramer-Rao bound to the probability distribution of the phases. It is:

$$\text{var}(z_0) = \sigma^2 \Big/ \frac{2}{N} \sum_{k=1}^{\Delta k} \left(\frac{2\pi k}{N}\right)^2 |X_k|^2, \tag{17}$$

where $k$ labels the harmonics in the digital Fourier transform (DFT) of the reference correlogram $I$:

$$I_j = \frac{1}{N} \sum_{k=0}^{N-1} X_k e^{2\pi i \frac{jk}{N}}, \tag{18}$$

$j = 1..N$, where $N$ is the number of measurement points; $i$ is the imaginary unit, $X_k$ are the complex harmonics' amplitudes, and $\Delta k$ is an effective spectrum summation interval, which, in practice, is much smaller, than the Nyquist limit $N/2$.

Let us examine how the variance (17) relates to the variance obtained by the CorCor method (7). Since continuous derivatives of the reference correlogram $I$ are needed in (7), we need to extent $I$ to the intermediate of net. We define $I(j+a)$, $0 \leq a < 1$, as summation of available Fourier harmonics, but include a corresponding phase shift

$$I(j+a) = \frac{1}{N} \sum_{k=0}^{N-1} Y_k e^{2\pi i \frac{jk}{N}};$$

$$Y_k = X_k e^{2\pi i \frac{ak}{N}}, \quad \text{at} \quad k \leq N/2; \tag{19}$$

$$Y_k = X_k e^{-2\pi i \frac{a(N-k)}{N}}, \quad \text{at} \quad k > N/2,$$

so that for $I(j+a)$ the necessary DFT symmetry property $Y_k = Y_{N-k}{}^*$ is retained; at the net knots $a = 0$ and (19) reduces to (18). The power spectrum of $I$ remains unchanged. Then the derivative is obtained as

$$\left.\frac{dI}{da}\right|_j = \frac{1}{N} \sum_{k=0}^{N-1} \left.\frac{dY_k}{da}\right|_{a=0} e^{2\pi i \frac{jk}{N}}.$$

Assuming that the derivative $dI/dz|_j$ in (7) is equal to $dI/da|_j$, and substituting it into the denominator of (7) which we call $L$, we obtain:

$$L = \sum_{j=1}^{N} \left(\frac{dI}{dz}\right)^2\bigg|_j = \sum_{j=1}^{N} \frac{dI}{dz}\bigg|_j \frac{dI^*}{dz}\bigg|_j = \frac{1}{N^2} \sum_{j=1}^{N} \sum_{\substack{k=0 \\ l=0}}^{N-1} \frac{dY_k}{da} \left(\frac{dY_j}{da}\right)^* \bigg|_{a=0} e^{2\pi i \frac{j(k-l)}{N}} =$$

$$\frac{1}{N} \sum_{\substack{k=0 \\ l=0}}^{N-1} \delta_{k,l} \frac{dY_k}{da} \left(\frac{dY_j}{da}\right)^* \bigg|_{a=0} = \frac{1}{N} \sum_{k=0}^{N-1} \left|\frac{dY_k}{da}\right|^2 \bigg|_{a=0}. \tag{20}$$

Here we have changed the order of summation and taken into account that the derivatives of $I$ are real values, and therefore squaring them is equal to multiplying with the conjugated value. In addition, we have used the Kronecker symbol $\delta_{k,l} = 0$, if $k \neq l$; $\delta_{k,k} = 1$ (not to be mistaken with the deviations) by means of which the orthonormality of Fourier components is expressed:

$$\delta_{k,l} = \frac{1}{N} \sum_{j=1}^{N} e^{2\pi i \frac{j(k-l)}{N}}.$$

Calculating the derivative of $Y_k$ according (19) we obtain

$$\left|\frac{dY_k}{da}\right|^2 = \left|\frac{dY_{N-k}}{da}\right|^2 = \left(\frac{2\pi k}{N}\right)^2 |X_k|^2 \quad at \quad k \leq N/2,$$

and, substituting it into (20), because the expression vanishes at k = 0, we finally find

$$L = \frac{2}{N}\sum_{k=1}^{N/2-1}\left|\frac{dY_k}{da}\right|^2 + \frac{1}{N}\left|\frac{dY_{N/2}}{da}\right|^2 = \frac{2}{N}\sum_{k=1}^{N/2-1}\left(\frac{2\pi k}{N}\right)^2 |X_k|^2 + \frac{(\pi)^2}{N}|X_{N/2}|^2. \quad (21)$$

This formula shows the equivalence of (7) and (17), if the Nyquist harmonics of the spectrum can be neglected, i.e. if $\Delta k \leq N/2-1$.

The equivalence goes down to the sparse nets containing just few z steps within the main correlogram oscillation period $\lambda_0$, but in still coarser discretizations it is lost. To demonstrate the equivalence of (7) and (17) we have considered the a correlogram of the form

$$I_i = \exp\left\{-\left[(j-j_0)/(2\lambda_0)\right]^2\right\} \times \cos\left[(2\pi/\lambda_0)(j-j_0)\right],$$

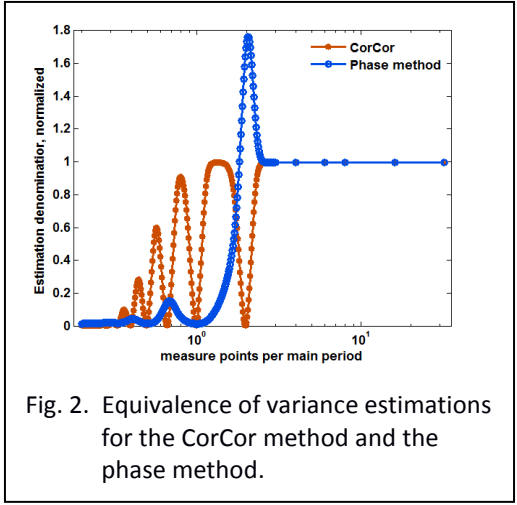

Fig. 2. Equivalence of variance estimations for the CorCor method and the phase method.

on a net which includes $z_0$. The derivatives appearing in (7) were taken analytically. Fig. 2 shows the comparison of the denominators of (7) and (17) for discretizations with different numbers of net points. Starting from 2.5 points per period both estimations are working correctly. At lower discretization point numbers they fail, since the correlogram spectra are not represented accurately. Yet, except the discretizations with knots near positions where the derivatives vanish, the estimation (7) is still working. This shows, that even for coarse correlograms the CorCor method gives a reasonable estimate for the surface height variance. In fact, the CorCor method can work even with just one point at the peak of the correlogram, but then (7) is not applicable, because it includes the zero derivative at the peak.

The equivalence of (7) and (17) for dense discretization nets is not surprising, because the Fourier transform of the cross-correlation function of the measured (J) and the reference (I) correlograms, the latter employed in the CorCor method, is equal to the multiplication of their Fourier transforms in the frequency domain. The phase term of the product contains phase differences of the measured and reference correlograms. The fitting of the so obtained phase distribution with a straight line using the method of least squares gives $z_0$ as the coefficient of the straight line's derivative, which is the very phase method in form of [16,17]. Nevertheless, the CorCor method is not identical to the phase method, and in accordance with the first subsection of this paper it is more accurate. This is because in reality the phase method contains additional noise-induced errors: First, it is subjected to errors due to the well-known 2π ambiguity. Second, the phase method does not take the spectral amplitude moduli in (17) from a reference correlogram (otherwise it would closely represent the CorCor method in the frequency domain) but from the measured correlogram itself. In this way the already mentioned loss of reference information takes place, and noise induced deviations of the $|X_k|$-weights used in the LS estimation of $z_0$ result in additional errors. The differences between $|X_k|_{ref}$ and $|X_k|_{meas}$ can easily be taken into account. Indeed, alternatively to way of [16,15], the estimation of the phase method's variance (17) can be obtained directly from the LS estimation of $z_0$ as follows. The correlogram phase distribution probability given in [16] after a slight renaming reads

$$p = \left(\frac{1}{\sqrt{2\pi}\sigma_k}\right)^N \prod_{k=1}^{\Delta k} \exp\left(-\frac{1}{2}\frac{\left(\phi_k + \frac{2\pi k}{N}z_0\right)^2}{\sigma_k^2}\right), \qquad (22)$$

where $\phi_k$ is the phase of $k^{th}$ harmonics in the Fourier expansion of measured correlogram (analogously to (18), but applied to $J$), $\sigma_k^2$ is variance of the $\phi_k$, the deviations supposed to be small. The maximization of (22) in accordance with the Maximum Likelihood Method results in LS formula for $z_0$ (a substantiation of the expression for $\sigma_k$ can be found in [16]):

$$\sigma_k = \sqrt{\frac{N}{2}}\frac{\sigma}{|X_k|};$$

$$\Phi = \sum_k \left(\phi_k + \frac{2\pi k}{N}z_0\right)^2 |X_k|^2 \to \min_{z_0} \;\Rightarrow\; \frac{\partial \Phi}{\partial z_0} = 0; \;\Rightarrow\; \sum_k 2\frac{2\pi k}{N}\left(\phi_k + \frac{2\pi k}{N}z_0\right)|X_k|^2 = 0; \;\Rightarrow\;$$

$$z_0 = \frac{\sum\limits_k |X_k|^2 \frac{2\pi k}{N} \phi_k}{\sum\limits_k |X_k|^2 \left(\frac{2\pi k}{N}\right)^2}. \qquad (23)$$

There are several possibilities to weight the points when performing LS fitting, but only the final formula of (23) results directly from (22) employing the Maximum Likelihood principle[1]. Now, the deviations of $z_0$ in (23) occurs, when the phases $\phi_k$ deviate from their true values; assuming the independence of deviations $\delta\phi_k$ [16] and applying the property of variance of several summed random variables, we obtain the variance of the *LS* estimation of $z_0$:

$$\sigma_{z_0}^2 = \frac{\sum\limits_m |X_m|^4 \left(\frac{2\pi m}{N}\right)^2 \sigma_m^2}{\left(\sum\limits_m |X_m|^2 \left(\frac{2\pi m}{N}\right)^2\right)^2} = \frac{\frac{N}{2}\sigma^2}{\sum\limits_m \left(\frac{2\pi m}{N}\right)^2 |X_m|^2}. \qquad (24)$$

This is once more the estimation (17), showing apropos that Cramer-Rao is efficient in this case as it is characteristic for normally distributed variates. Let us stress that (24) is exact only if one uses the spectrum amplitudes of reference correlogram to weight the phases. This emphasizes the importance of the use of reference information. In today praxis, however, one uses the spectral amplitudes $|X_k|_{meas}$ of the current measured correlogram $J$ for the weighting (if weights at all) instead of that of the system-characteristic $|X_k|_{ref}$ of $I$ and then, because of $\sigma^2_{|Xk|meas} = N\sigma^2$ (cf. [16]), variance of the $z_0(\phi_k, |X_k|_{meas})$ becomes

$$\sigma_{z_0}^2 = \sum_m \left(\frac{\partial z_0\left(\varphi_m, |X_m|_{meas}\right)}{\partial \varphi_m}\right)^2 \sigma_m^2 + \sum_m \left(\frac{\partial z_0\left(\varphi_m, |X_m|_{meas}\right)}{\partial |X_m|_{meas}}\right)^2 \sigma^2_{|X_m|_{meas}} =$$

$$\frac{\frac{N}{2}\sigma^2}{\sum\limits_m \left(\frac{2\pi m}{N}\right)^2 |X_m|^2} + N\sigma^2 \sum_m \left(\frac{\partial z_0\left(\varphi_m, |X_m|_{meas}\right)}{\partial |X_m|_{meas}}\right)^2. \qquad (25)$$

---

[1] Refering to [15] we might remark that the formulas (27) and (38) of this paper do not result from the formula (22) as it is implied in [15].

Comparing to (24), in (25) there is an additional second summand; using the simulation we show below that this term can be relevant compared to the first summand. This means that in praxis the phase method is still less accurate, than the CorCor.

## 6. Simulation of height evaluation with the compared methods

A simulation has been performed employing correlograms in form

$$J_j = \exp\left\{-\log(2)\left[(z_j - z_0)/W\right]^2\right\} \times \cos\left[(2\pi/\lambda_0)(z_j - z_0)\right] + \sigma \cdot rand(j), \tag{26}$$

where the correlogram amplitude is put to unit; $z_0$ is the position of the correlogram maximum representing the surface height position, $W$ is the half-width of the correlogram, $\lambda_0$ is the wavelength of its main oscillation, $\sigma$ is the noise dispersion, $rand$ is the function producing normally-distributed random values with zero mean value and the variance of unity. This means that white noise has been used to represent the noise of measurement. The discretization step has been here also chosen as the length unit, so $z_j = j$; the whole scan range is $N = 1024$. If not specified otherwise, $W = 2\lambda_0$, and $\lambda_0 = 8$. The correlogram $J$ at a noise level of $\sigma = 0.1$, its amplitude and phase spectra can be seen in Fig. 1 of Supplementary Materials. Noises with relative levels higher than 0.1 have not been simulated because some of the considered methods do not converge for such noises. For all the variants of the phase method the fringe order has been fixed to be correct, in this way the simulation results are cleared from the $2\pi$ jump errors. Every noise-level-point of the simulation results represents statistics of 1 000 repetitions, increasing this number does not produce any distinguishable difference.

Fig. 3 and Fig. 4 show the outcome of simulations. First of all, we notice that in agreement with the conclusion of subsection 1. the surface height estimation error of the CorCor procedure is the lowest among the considered methods.

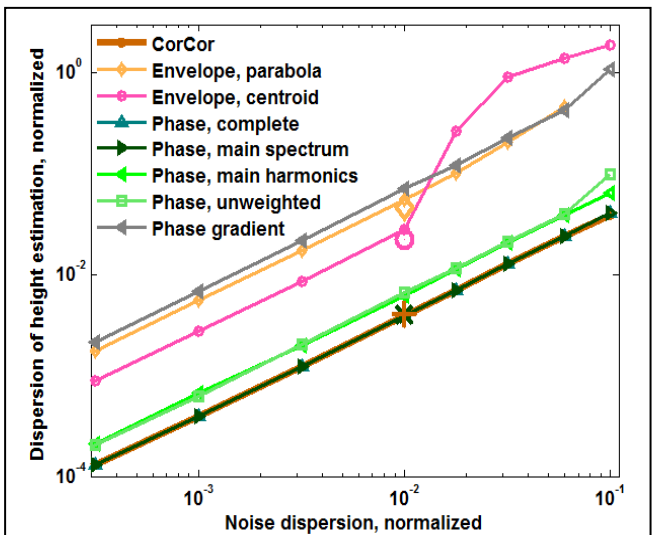

Fig. 3. Variance of height evaluation for the compared methods employing spectrum of the reference correlogram. Large symbols at $\sigma = 0.01$ represent estimations (7), (13), (16), and (17).

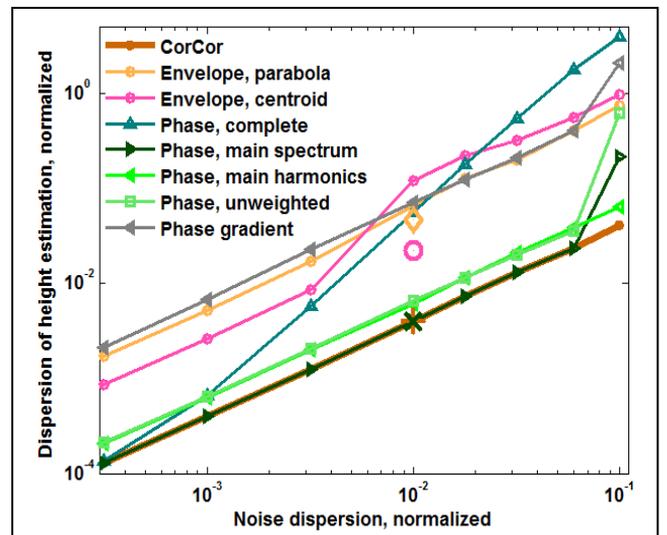

Fig. 4. Variance of height evaluation for the compared methods employing spectrum of the current correlogram. Large symbols represent estimations (7), (13), (16), and (17).

In accordance with the results of subsection 5., we observe the coincidence of the error curves of the CorCor method and two modifications of the phase method in the Fig. 3. However, even here at higher noise levels the CorCor method is superior (see Supplementary materials Fig. 2). This does not contradict to the point of (21), because the estimation (17) holds for small noise deviations only. We want to mention once more that the 2π jumps, which are characteristic for the phase methods especially at elevated noise levels, were deliberately excluded from the data of Figs. 3 and 4.

According to (7), (13), (16), and (17) the variance of height estimation is proportional to the noise dispersion. Hence, all simulated curves in Figs. 3 and 4 should be straight lines with a pitch of one decade per one decade. This is true for all curves at low noises, which by the way indicates absence of systematic errors in the simulation. At higher noise levels only the curves representing the CorCor procedure and the monochromatic phase evaluation (called "main harmonics" in the figures) retain this property. The variance of height estimation by all other methods tends to grow faster at higher noise levels – the curves deflect upwards. Moreover, only the CorCor straight line and two of the phase method straights lines reside on the levels predicted by the analytical estimations. As far as the envelop methods are concerned, the estimation for the centroid method (16) only gives a lower bound that does not account for the correlations of neighboring envelope points, therefore the corresponding straight line is located slightly above the circle representing the result (16) at a noise dispersion of 0.01. The same is true for formula (13) giving the variance of the envelope parabola fitting method. Eq. (13) is derived from the maximum likelihood principle and a probability estimation analogous to (2), which is exact only for uncorrelated variates.  All in all however, the discrepancies between analytics and the simulation are small for the envelope methods too, indicating a good applicability of the estimations (13) and (16). Finally, we notice without discussing in detail that the simulated variance of the centroid method rises abruptly at noise levels above 0.01.

Let us now address the performance of the phase method implementations. In this simulation we introduced a threshold of 5% of the maximal spectral amplitude to select the main set of harmonics of the spectrum. In Figs. 3 and 4 the curve named "Phase, complete" represents the weighted summation over all "physical" harmonics 1..N/2-1 according to (23); the curve named "Phase, main spectrum" represents the weighted summation over the specified main set of harmonics.

Mostly, the accuracy of the phase methods is higher, than that of the envelope methods, which is widely known from practice. The phase gradients in the phase gradient method were calculated by taking the finite phase differences on the net and averaged without weighting. The corresponding lines in the Figs 3. and 4. are located above the others, thus, the phase gradient method is less precise, which is also known from practice [16]. Often, this procedure is used only to ascertain the fringe order [4]. This is because the phase gradient method [17] is not subjected to 2π errors, since only phase differences of the neighbor harmonics are involved. The version of the phase method following (23) gives - when properly weighted phases of selected harmonics range are used - an accuracy almost equal to that of the CorCor method, although some deviation at higher noises are present. There exist two other variants of phase evaluation – an unweighted summation and the use of only one main harmonics. The latter method is inferior to a summation following (23), which is in agreement with the variance estimation (17), where the resulting variance decreases the more harmonics are summed. Then, it is due to the fact that the weights of harmonics lying outside the main set are negligible that the "complete" and "main spectrum" curves coincide in Fig. 3. The curve of the method's variant using unweighted main spectrum harmonics coincides with the straight line of the one harmonics variant, because the height estimations from any of the main set harmonics are very close to that of the very main harmonics of the highest amplitude, and they are just averaged by the summation. Still, at high noise levels harmonics with small amplitudes show higher phase deviations compared to the main harmonics (contrary to the result [16], which however holds only for small noise). Therefore, being averaged without weighting, the main spectrum harmonics give a little higher variance, than the main harmonics itself.

The good agreement between the estimations (7), (13), (16), and (17) and simulations confirms the correctness of both the analytic expressions and the simulations. Moreover, their exact coincidence in the case of CorCor and the optimal phase method shows that the Cramer-Rao estimation is efficient for both.

The whole dataset of Fig. 3 is obtained using, in one way or another, the reference correlogram, which is not disturbed by noise. In the implementations of the phase method it has been employed for weighting and to ascertain the frequency interval containing the main spectrum. In the envelope methods it has been used to find the half-envelope boundaries. The CorCor method harvests the complete information of reference correlogram. Once more, we stress that in praxis the methods are used without taking any advantage of reference correlogram's information, thus, being more self-contained, they are less accurate. To show this we have performed a simulation of the mentioned methods when not the reference correlogram, but instead the currently measured correlogram is used. The results are shown in the Fig. 4. Here the CorCor curve is plotted solely to have a bench mark – this method employs a reference correlogram anyway. A comparison with Fig. 3 shows that at higher noise levels the method's variances are higher, when no reference correlogram is used. This behavior is most pronounced for the phase method which uses all harmonics, and this is due to deviations of the weights, which in their turn are induced by deviations within the spectrum. At a certain noise level the method's $z_0$-variance even rises above that of all the other methods. Some weights, being negligible in the basic reference spectrum, and corresponding to harmonics located outside of the main spectral interval, can randomly obtain a discernible values, giving rise to strongly deviating phase when summed (see the Supplementary materials Fig. 1). But even the most accurate "main spectrum" phase technique deflects from the CorCor line. The envelope method variances are not anymore following the predictions (13) and (16) at the noise level of 0.01, but lie above them, having deflected from the basic straight lines earlier. Without the use of the reference correlogram the variances of the envelope methods are now influenced by the uncertainty of the half-envelope height boundaries.

Finally, it is useful to take a look on the dependence of the $z_0$ variance on the width of the correlogram packet. In the following simulation we have kept $\lambda_0 = 8$, but varied the packet half-width $W$. Throughout the self-contained approach has been applied - the reference correlogram information has been neglected. The results of the simulation are shown in Fig. 5. They substantiate the common assumption that for the phase methods employing the main spectrum the height estimation becomes more accurate with increasing packet width. Although not obvious, this also happens to be the case for the CorCor method. Indeed, according to (7) it is not the concrete form of $I$ that is essential for the value of the variance, but the steepness of the gradients of $I$ and the number of measurement points building up the packet. This number rises and, although the gradients remain unchanged, the accuracy increases. In contrast, the accuracy of the envelope methods decreases despite the growing number of fitting points, as the steepness of the envelope wings becomes less (cf. (13)). Turning to the phase method which uses all the available harmonics we notice the decreasing precision. With the growing packet width the number/weights of main-spectrum harmonics reduce, the straight line fitting the phase spectrum (23) becomes less stable to the disturbances of outsider harmonics. This occurs despite the growing precision of the main phase points, since for the fitting-line stability the number/weights of dependable points are essential. In other terms, with the growing packet width the second summand in (25) grows and is dominant over the first one. The growing disturbance of the fitting line is in effect for the phase gradient method too.  This result is in conflict with the wide-spread opinion that a growing packet width results in higher precision of the phase methods in general. The behavior differs for the different method implementations.

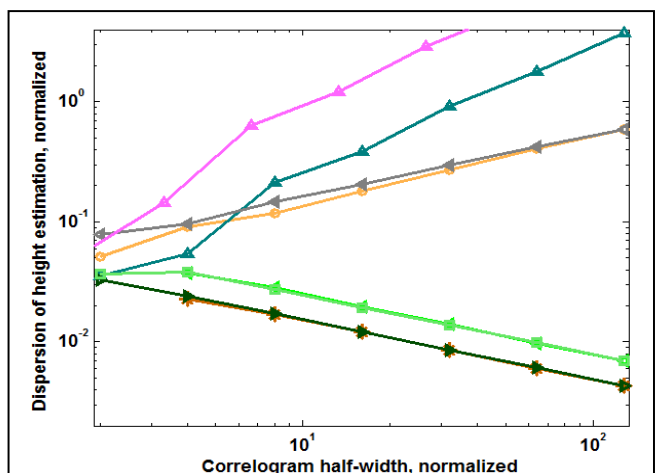

Fig. 5. Variance of height evaluation in dependence on the correlogram packet width (given in discretization units). The coloring of the lines is the same, as in Figs 2. and 3.

## 7. Choosing of the reference correlogram

The question of how to obtain the reference correlogram in practice can overshadow the attraction of the CorCor method despite of all its advantages. It is essential to mention that if the reference correlogram contains noise of standard deviation $\sigma_{ref}$, the deviations of the CorCor estimations of height only increase by a factor of $sqrt(\sigma^2 +\sigma_{ref}^2)$. Thus, the expected worsening is limited to a factor of sqrt(2) even for the worst possible procedure of reference selection. Choosing correlograms with the highest contrast and, possibly, averaging them, reduces the worsening to an even lower level.

The reference correlogram can be selected/produced in three following ways which are listed in the order of decreasing quality:
- by measuring and averaging of correlograms with the same WLI device on a smooth horizontal surface of the material in question without light disturbing particularities;
- by the selection of several correlograms with highest contrast among the pixel-correlograms obtained in the same current measurement  and their height-shift-independent averaging;
- by selecting one of the correlograms with the highest contrast as the reference correlogram.

It is also theoretically possible, but hardly reasonable in practice, to synthesize the reference correlogram employing known devise and surface characteristics.

In the following instances we have used the simple third variant of the above list.

## 8. Practical demonstration of the accuracy of the CorCor method

 So far the CorCor method was validated by analytical and simulation approaches. Here we give two examples illustrating the exceeding practical effectiveness of this method in the areas, where envelop and phase methods are correspondingly known to be the most appropriate. It is established that envelope methods work properly on the rough surfaces, where the phase methods fail owing to the $2\pi$ uncertainty, and the phase methods demonstrate an eminent accuracy on smooth surfaces.

Fig. 6 shows a rough 9 μm groove in a steel surface evaluated with the envelope and CorCor methods. With the envelop method a pixel correlogram is classified as unsuitable for evaluation, if the maximal contrast is less than 5% of the hardware-determined intensity maximum; the level of 5% has been established during many years of practice as the lowest to screen out outliers. With the CorCor method a pixel is sorted out if the deviation from the neighboring pixel's values exceeds three height variances calculated on horizontal surface areas, this selection also is meant to eliminate outliers. With this thresholding both methods tolerate a roughness comparable to that of the horizontal surface, but not stronger. Obviously, fraction of the pixels screened out during the CorCor procedure is much smaller: the CorCor method is much more robust. The superiority of the CorCor method for the case of a smooth surface is demonstrated in Fig. 7. Here, the height estimation of the CorCor method along a trace line on a smooth Si surface is compared with that of a standard implementation of phase method [18]. Actually, in this case the CorCor method has provided the amazingly low variance of the surface height of 0.17 nm. This two examples give experimental illustration and confirmation of the conclusion of subsection 1. that the CorCor method is destined to be the most stable and accurate in the presence of uncorrelated noise.

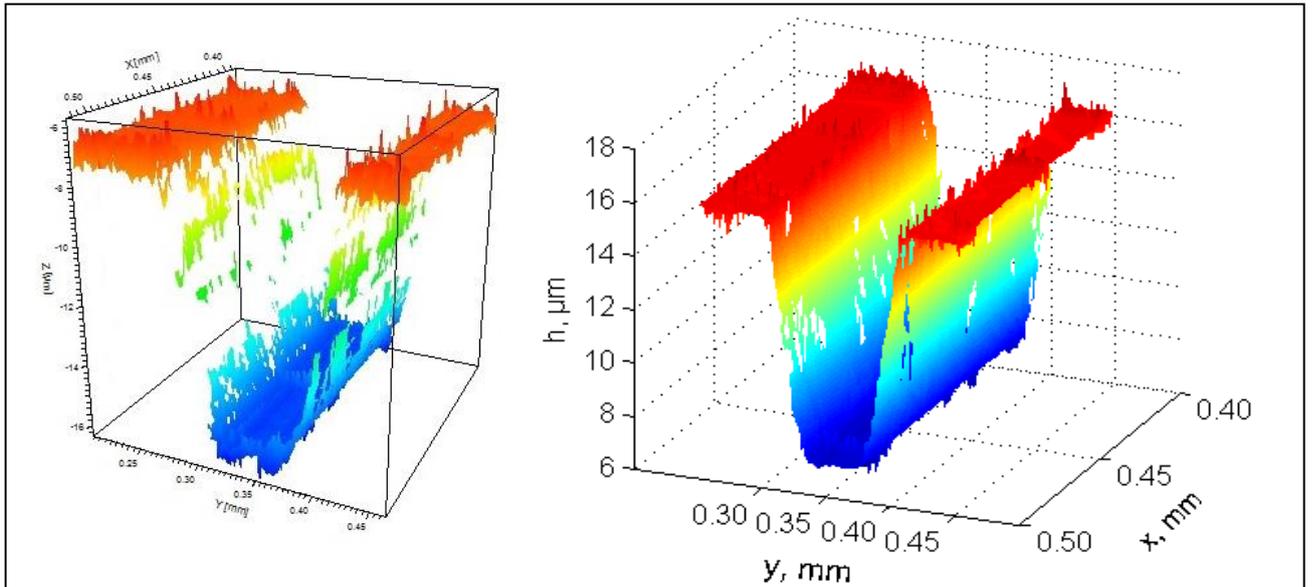

Fig. 6. Comparison of a rough groove surface evaluated by the standard half-envelop-parabola (left) and CorCor methods. Pixels which cannot be evaluated with the corresponding method are not colored.

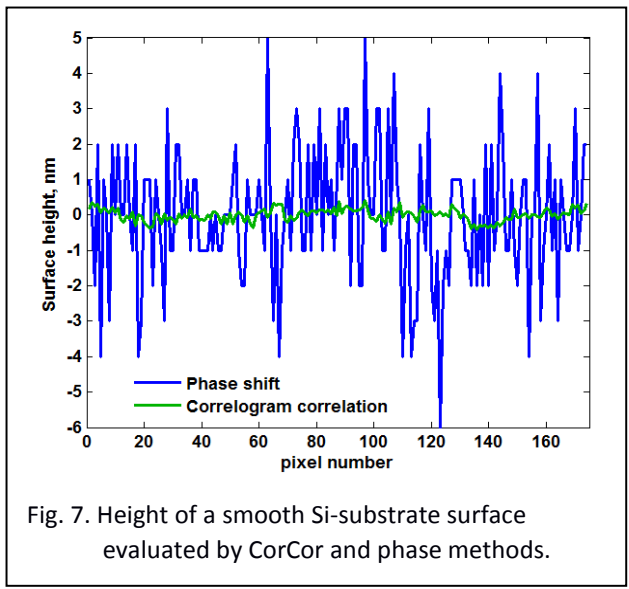

Fig. 7. Height of a smooth Si-substrate surface evaluated by CorCor and phase methods.

## 9. Discussion

In the practice of WLI there exist situations, where the correlograms obviously loose the similarity of their shape to that of any possible reference correlogram. This situations can be divided into three classes as follows: i) The measured correlograms depend on the pixel position owing to properties of the investigated surface. ii) The appearing correlograms are sums of several different correlograms. As is shown in [6], correlograms appearing due to reflection from different surfaces sum themselves additively, even if the coherence zones overlap and the reflected rays interfere. Thus, they are actually separable in the sense of linear algebra. A typical instance of this summation are correlograms received from thin films, where reflections from two film interfaces result in a sum of two correlograms. Another instance is reflection in the vicinity of a sharp step, where due to the limited resolution of the interferometer optics some pixels of the recording camera receive correlograms from surfaces of both levels. Similar correlograms are often obtained at pores or scratches in a smooth surface. iii) The correlograms at different surface heights differ in shape – the spectrum change due to height is not just a linear phase shift, as it is usually characteristic for the WLI.

In case i), whether the CorCor method can be applied dependends on the number of possible correlogram types. If limited, the procedure can be performed consequently using the set of possible reference correlograms with a following selection of the match with the best correlation. Otherwise, non-pattern-specific methods – envelop or phase techniques - have to be employed. This is also true in case iii), unless a shape determination employing correlations with a variety of possible reference correlograms can be used to ascertain the correlogram location.

In case ii) several correlograms have to be combined in a way similar to that indicated in [6], [7]. The procedure is not anymore just the plain search for maximal cross-correlation, but the basic idea to fit the measured correlogram to a reference one is still instrumental.

Obviously, involving the available reference information is always useful, unless the information is irrelevant, which has to be checked in view of any concrete application.

The CorCor procedure is not extensively time consuming, because the calculation of the cross-correlation function is done by a couple of direct and inverse Fast Fourier Transforms. It can be further quickened by a complete transfer into the frequency domain.

## Summary

It is shown that the correlogram correlation method is equivalent to the direct application of maximal likelihood criterion, i.e. that the surface height estimated using this method is correct with maximal probability. The only condition needed to reach this conclusion is that the noise at different scan axis points is uncorrelated. The statement is substantiated by deriving and comparing analytical expressions for the variances of heights estimated with the common envelope and phase methods, as well as by simulations. The obtained variance estimations are of interest on their own, e.g. for comparing the performance of conventional methods. Besides its high accuracy, the correlogram correlation method provides a simple solution to the widespread problem of how to decide whether a pixels correlogram is suited for the evaluation, the criterion being a high value of covariance with a reference correlogram.

## Acknowledgement

This research is supported by Federal Ministry for Economic Affairs and Energy on the basis of a decision by the German Bundestag under the project ID EP160748.

**Supplementary Materials**

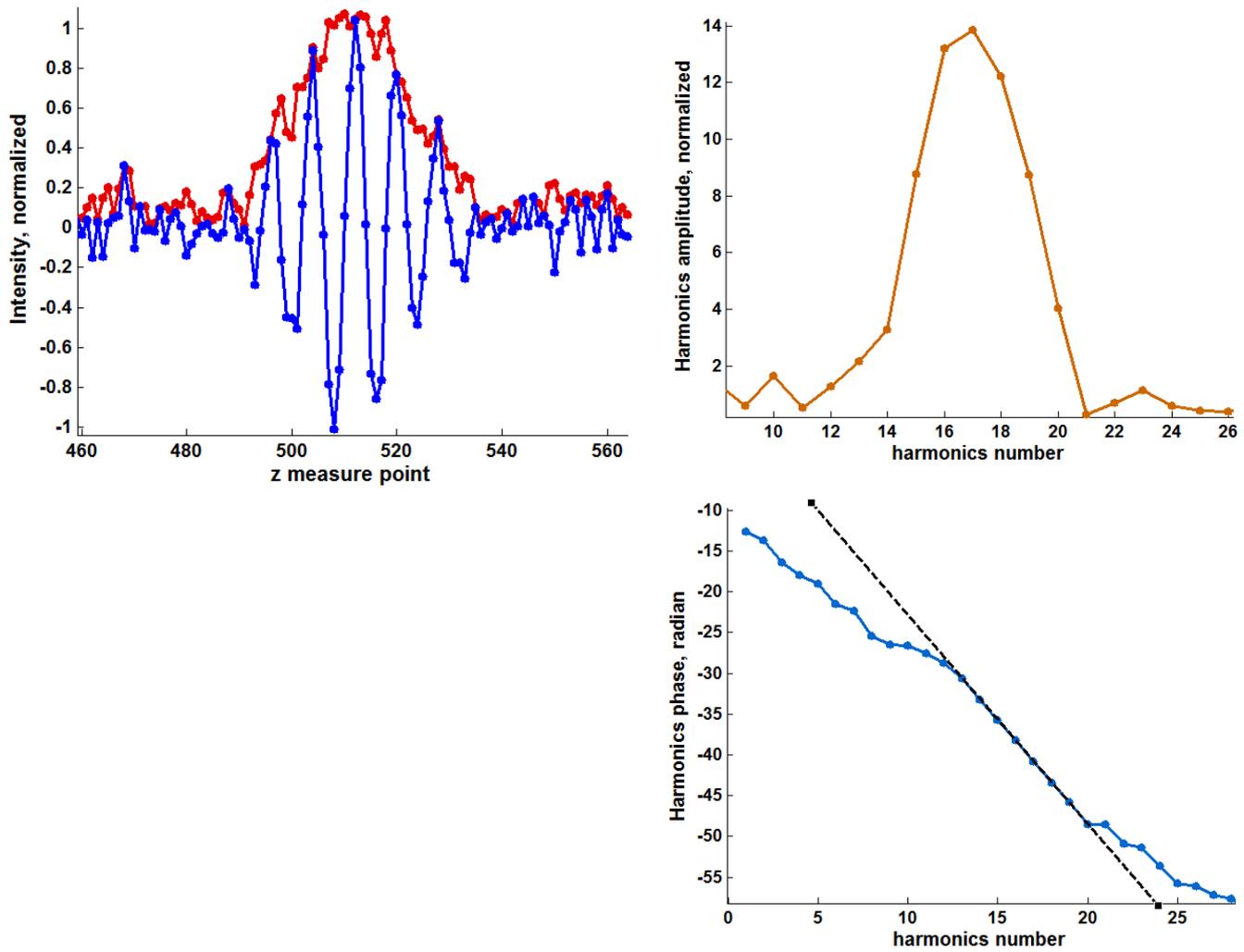

Fig. 1. Synthesized correlogram, its unfiltered envelope (red line), its amplitude and phase spectra. An instance at the noise of dispersion 0.1.

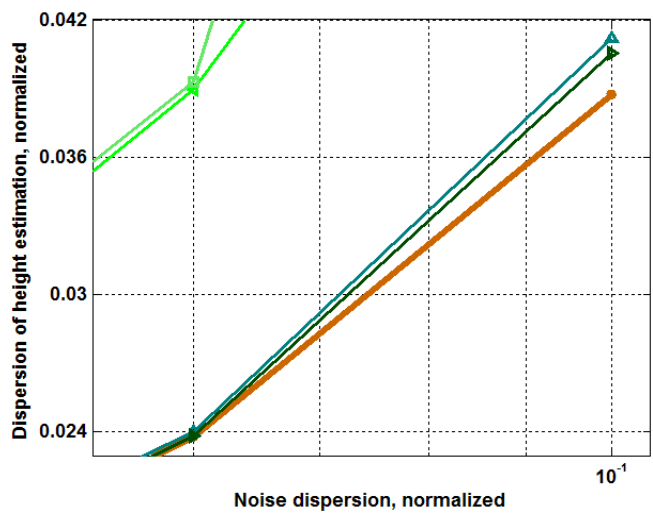

Fig. 2. Zoomed excerpt from the Fig. 3 of the main text.